\documentclass[
showpacs,preprintnumbers,amsmath,amssymb]{revtex4}
\usepackage{amsmath}
\usepackage{graphicx}
\usepackage{subfig}
\usepackage{hyperref}
\usepackage{amssymb}
\usepackage[english]{babel}
\usepackage{epsfig}
\usepackage{wasysym}
\usepackage{graphicx}
\usepackage{indentfirst}
\usepackage{amsmath}
\usepackage{amsfonts}
\usepackage{amssymb}
\usepackage{enumerate}

\begin{document}

\newcommand{\vecbo}[1]{\mbox{\boldmath $#1$}}
\newtheorem{theorem}{Theorem}
\newtheorem{acknowledgement}[theorem]{Acknowledgement}
\newtheorem{algorithm}[theorem]{Algorithm}
\newtheorem{axiom}[theorem]{Axiom}
\newtheorem{claim}[theorem]{Claim}
\newtheorem{conclusion}[theorem]{Conclusion}
\newtheorem{condition}[theorem]{Condition}
\newtheorem{conjecture}[theorem]{Conjecture}
\newtheorem{corollary}[theorem]{Corollary}
\newtheorem{criterion}[theorem]{Criterion}
\newtheorem{definition}[theorem]{Definition}
\newtheorem{example}[theorem]{Example}
\newtheorem{exercise}[theorem]{Exercise}
\newtheorem{lemma}[theorem]{Lemma}
\newtheorem{notation}[theorem]{Notation}
\newtheorem{problem}[theorem]{Problem}
\newtheorem{proposition}[theorem]{Proposition}
\newtheorem{remark}[theorem]{Remark}
\newtheorem{solution}[theorem]{Solution}
\newtheorem{summary}[theorem]{Summary}
\newenvironment{proof}[1][Proof]{\textbf{#1.} }{\ \rule{0.5em}{0.5em}}
\hypersetup{colorlinks,citecolor=green,filecolor=magenta,linkcolor=red,urlcolor=cyan,pdftex}

\newcommand{\be}{\begin{equation}}
\newcommand{\ee}{\end{equation}}
\newcommand{\bea}{\begin{eqnarray}}
\newcommand{\eea}{\end{eqnarray}}
\newcommand{\beaa}{\begin{eqnarray*}}
\newcommand{\eeaa}{\end{eqnarray*}}
\newcommand{\Lhat}{\widehat{\mathcal{L}}}
\newcommand{\nn}{\nonumber \\}
\newcommand{\e}{{\rm e}}


\title{Cosmological viable $f(R,T)$ dark energy model: dynamics and stability}
\author{E. H. Baffou$^{(a)}$\footnote{e-mail:baffouh.etienne@yahoo.fr}, A. V. Kpadonou$^{(a,c)}$\footnote{e-mail: vkpadonou@gmail.com}, M. E. Rodrigues$^{(b)}$\footnote{e-mail: esialg@gmail.com},  
M. J. S. Houndjo$^{(a,d)}$\footnote{e-mail:
sthoundjo@yahoo.fr} and J. Tossa$^{(a)}$\footnote{e-mail: joel.tossa@imsp-uac.org} }
\affiliation{$^a$ \, Institut de Math\'{e}matiques et de Sciences Physiques (IMSP), 01 BP 613,  Porto-Novo, B\'{e}nin\\
$^{b}$\,Faculdade de Ci\^encias Exatas e Tecnologia, Universidade Federal do Par\'a - Campus Universit\'ario de Abaetetuba, CEP 68440-000, Abaetetuba, Par\'a, Brazil\\
$^{c}$\, Ecole Normale Sup\'erieure de Natitingou - Universit\'e de Parakou - B\'enin\\ 
$^{d}$\, Facult\'e des Sciences et Techniques de Natitingou - Universit\'e de Parakou - B\'enin }

\begin{abstract}
In this paper we undertake the modified theory of gravity $f(R,T)$, where $R$ and $T$ are the Ricci scalar and the trace of the energy momentum tensor, respectively. Imposing the conservation of the energy momentum tensor, we obtain a model about what dynamics and stability are studied. The stability is developed using the de Sitter and power-law solutions. The results show that the model presents stability for both the  de Sitter and power-law solutions. Regarding the dynamics, cosmological solutions are obtained by integrating the background equations for both the low-redshift and High-redshift regimes and are consistent with the observational data.
\end{abstract}
\maketitle 

\section{Introduction}
Nowadays, modifying the law of gravity is a possible way to explain the acceleration mechanism of the universe \cite{1dediego,2dediego}. Various theories are developed to explain some characteristics and properties of the dark energy, known as the responsible of the expanded acceleration  of the universe, but it is not clear which king of modified theory of gravity will finally prevail and attention has to be attached to each one. \par
This paper is devoted to the some cosmological studies in the so-called $f(R,T)$ theory of gravity. This theory takes its origin from the fact the cosmological constant may be taken as a trace{\footnote{The trace here means the one of the energy momentum tensor.}} dependent function, the so-called  ``$\Lambda(T)$ gravity" in order to guarantee the interaction between the DE and the ordinary content of the universe, and this is favoured by the recent cosmological data \cite{lambdadeT}. This later has been extended viewing the algebraic function that characterizes the Lagrangian density as functions of both the Ricci $R$ scalar and the trace $T$ of the energy momentum tensor, namely, $f(R,T)$ \cite{frtoriginal}. The dependence on $T$ may be a consequence of the universe being partially filled by an exotic imperfect fluid, or consequence of quantum effects coming from conformal anomaly. Several works have been developed in the framework of this kind of modified theory of gravity and considerable results have 
been found \cite{lesfrt}-\cite{anil}.  However, anywhere, these theories took into account the important aspect of guaranteeing the conservation of the energy momentum tensor. This feature has been first undertaken by Alvarenga and collaborators \cite{papierdiego} where they consistently ensured the conservation of the energy momentum tensor, from which they constructed a $f(R,T)$ model. In that paper, they investigate the dynamics of scalar perturbation within the obtained model and focused they attention to the sub-Hubble modes, and showed that through the quasi-static approximation the results are very different from the ones derived in the frame of the concordance $\Lambda CDM$ model, constraining of the validity of this kind of model. \par 
The result obtained in that paper is quite reasonable due to the choice of the ordinary matter content and the determination of the integration constant. These factors strongly influenced the result in that paper and as our goal in this paper, we propose to keep the ordinary matter, not only as the non-relativistic one as performed in \cite{papierdiego}, but as a mixture 
of non-relativistic matter (dust) and  relativistic matter (radiation), despite the current low proportion of this latter.  Rather than studying the dynamics of scalar perturbations, we will focus our attention to the cosmological dynamic in the low-redshift and high-redshift regimes. Moreover the stability of de Sitter and power-low solutions within the model under consideration.\par
The paper is organized as follows: in Sec. \ref{section2} we construct the model consistent with the vanishing divergence of the energy momentum tensor. The stability of the critical points of the dynamics system are checked in the Sec. \ref{section3} and the one of the perturbation functions within the model under consideration is developed in the Sec. \ref{section4}. The Sec. \ref{section5} is devoted to the study of the cosmological dynamics with the considered model. The conclusion is presented in the Sec. \ref{section6}.

\section{Obtaining the model according to energy-momentum tensor conservation}\label{section2}

We start this work writing the action in the following form 
\begin{eqnarray}
S=\frac{1}{2\kappa}\int \sqrt{-g}\left[f(R,T)+\mathcal{L}_m\right]\,\,,\label{eti1}
\end{eqnarray}
where $R$, $T$ are the curvature scalar and the trace of the energy momentum tensor, respectively, and $\kappa^2=8\pi G$, $G$ being the gravitation constant. The energy momentum tensor is defined from the matter Lagrangian density $\mathcal{L}_m$ by
\begin{eqnarray}
T_{\mu\nu}=-\frac{2}{\sqrt{-g}}\frac{\delta\left(\sqrt{-g}\mathcal{L}_m\right)}{\delta g^{\mu\nu}}.\label{eti2}
\end{eqnarray} 
By varying the action with respect to the metric $g^{\mu\nu}$, one gets the general equations of motion
\begin{eqnarray}
f_{R}R_{\mu\nu}-\frac{1}{2} g_{\mu\nu}f(R,T)+(g_{\mu\nu}\Box-\nabla_{\mu}\nabla_{\nu})f_{R}= \kappa^{2}T_{\mu\nu}-
f_{T}(T_{\mu\nu}+\Theta_{\mu\nu})\,,\label{eti3}
\end{eqnarray}
where $\Theta_{\mu\nu}$ is determined by 
\begin{eqnarray}
\Theta_{\mu\nu}\equiv g^{\alpha\beta}\frac{\delta T_{\alpha \beta}}{\delta g^{\mu\nu}}=-2T_{\mu\nu}+g_{\mu\nu}\mathcal{L}
-2g^{\alpha\beta}\frac{\partial^2 \mathcal{L}_m}{\partial g^{\mu\nu}\partial g^{\alpha \beta}}\label{eti4}.
\end{eqnarray}
In order to reach the expression of the covariant derivative of the energy-momentum tensor and extract the one of the algebraic function, one perform the covariant derivative of (\ref{eti4}), as
\begin{eqnarray}
\nabla^{\mu}\big[ f_{R}R_{\mu\nu}-\frac{1}{2} g_{\mu\nu}f(R,T)+(g_{\mu\nu}\Box-\nabla_{\mu}\nabla_{\nu})f_{R}\big]
=\nabla^{\mu}\big[k^{2}T_{\mu\nu}-
f_{T}(T_{\mu\nu}+\Theta_{\mu\nu})\big] ,\label{eti5}
\end{eqnarray}
which can be rewritten as
\begin{eqnarray}
f_{R}\nabla^{\mu}R_{\mu\nu}+
R_{\mu\nu}\nabla^{\mu}f_{R}-\frac{1}{2}g_{\mu\nu}\left( f_R\nabla^\mu R+f_T\nabla^{\mu}T \right)+
(g_{\mu\nu}\nabla^{\mu}\Box-
\nabla^{\mu}\nabla_{\mu}\nabla_{\nu})f_{R}\nonumber\\
= k^{2}\nabla^{\mu}T_{\mu\nu}-
f_{T}(\nabla^{\mu}T_{\mu\nu}+\nabla^{\mu}\Theta_{\mu\nu})
-(T_{\mu\nu}+\Theta_{\mu\nu})\nabla^{\mu}f_{T} .\label{eti6}
\end{eqnarray}
By evaluating the forth term of the above expression, one gets 
\begin{eqnarray}
\left(g_{\mu\nu}\nabla^{\mu}\Box-\nabla^{\mu}\nabla_{\mu} \nabla_{\nu}\right)f_{R}&=&(\nabla_{\nu}\Box-\Box \nabla_{\nu})f_{R} \\ \nonumber
&=& -R_{\mu\nu}\nabla^{\mu}f_{R}\,\,.
\end{eqnarray}
From the above equation, one gets the following expression
\begin{eqnarray}
\nabla^{\mu}T_{\mu\nu}=\frac{1}{\kappa^2-f_T}\left[f_T\nabla^{\mu}\Theta_{\mu\nu}+\left(T_{\mu\nu}
+\Theta_{\mu\nu}\right)\nabla^{\mu}f_T-\frac{1}{2}g_{\mu\nu}f_{T}\nabla^{\mu}T\right] .\label{eti8}
\end{eqnarray}
By assuming that the matter content of the universe is a perfect fluid, one can write the energy momentum tensor as
\begin{eqnarray}
T_{\mu\nu}=\left(\rho+p\right)u_{\mu}u_{\nu}-pg_{\mu\nu},
\end{eqnarray}
where $\rho$ and $p$ are the energy density and the pressure of the ordinary matter, respectively, $u^{\mu}$ is the four-velocity such that $u^{\mu}u_{\mu}=1$. Therefore, the Lagrangian density may be chosen as $\mathcal{L}_m=-p$, and the tensor $\Theta_{\mu\nu}=-2T_{\mu\nu}-pg_{\mu\nu}$. Then, after some elementary transformations, Eq.~(\ref{eti8}) takes the following expression
\begin{eqnarray}
\nabla_{\mu}T^{\mu}_{\nu}=\frac{1}{\kappa^2+f_T}\left\{-\omega\delta^{\mu}_{\nu}f_T
\partial_{\mu}\rho+\left(1-3\omega\right)\left[
\left(-T^{\mu}_{\nu}-\omega \rho \delta^{\mu}_{\nu}
\right)f_{TT}-\frac{1}{2}\delta^{\mu}_{\nu}f_T\right]
\partial_{\mu}\rho\right\}\,,
\end{eqnarray} 
where we used the barotropic equation of state $p=\omega \rho$. By setting $\nu=0$,  one gets
\begin{eqnarray}
\dot{\rho}+3H\rho\left(1+\omega\right)=-\frac{\dot{\rho}}{\kappa^2+f_T}\left\{\omega f_T+\left(1-3\omega\right)\left[\rho\left(1+\omega\right)
f_{TT}+\frac{1}{2}f_T\right]\right\}.\label{eti11}
\end{eqnarray}
In order to ensure a null divergence of the energy momentum tensor, one has to vanish the r.h.s of Eq.~(\ref{eti11}), leading to the differential equation
\begin{eqnarray}
2\left(1+\omega\right)Tf_{TT}-
\left(\omega-1\right)f_T=0\,\,,
\end{eqnarray}  
whose general solution reads
\begin{eqnarray}
f(T)=\alpha T^{\frac{1+3\omega}{2(1+\omega)}}+\gamma\,\,,
\end{eqnarray}
where $\alpha$ and $\gamma$ are integration constants. In what follows, we assume $\gamma=0$ and search for $\alpha$ through initial conditions. We then assume that at the present time $t_0$ the $f(R,T)$  model recovers the $\Lambda CDM$ one, i.e., $f(R_0,T_0)=R_0-2\Lambda$. In this paper, we propose to work will the model $R+f(T)$. Therefore, one gets the algebraic function $f(T)$ as 
\begin{eqnarray}
f(T)=-2\Lambda\left(\frac{T}{T_0}\right)^{\frac{1+3\omega}{2(1+\omega)}},
\end{eqnarray}
meaning that the constant $\alpha$ depends on the cosmological constant $\Lambda$, the parameter of ordinary equation of state $\omega$ and the current trace of the energy momentum tensor $T_0$ as
\begin{eqnarray}
\alpha=-2\Lambda T_0^{-\frac{1+3\omega}{2(1+\omega)}}.\label{eti15}
\end{eqnarray}

\section{ Studying the stability of the critical point of the dynamic system}\label{section3}

\begin{figure}[h]
\centering
\begin{tabular}{rl}
\includegraphics[height=7cm,width=7cm]{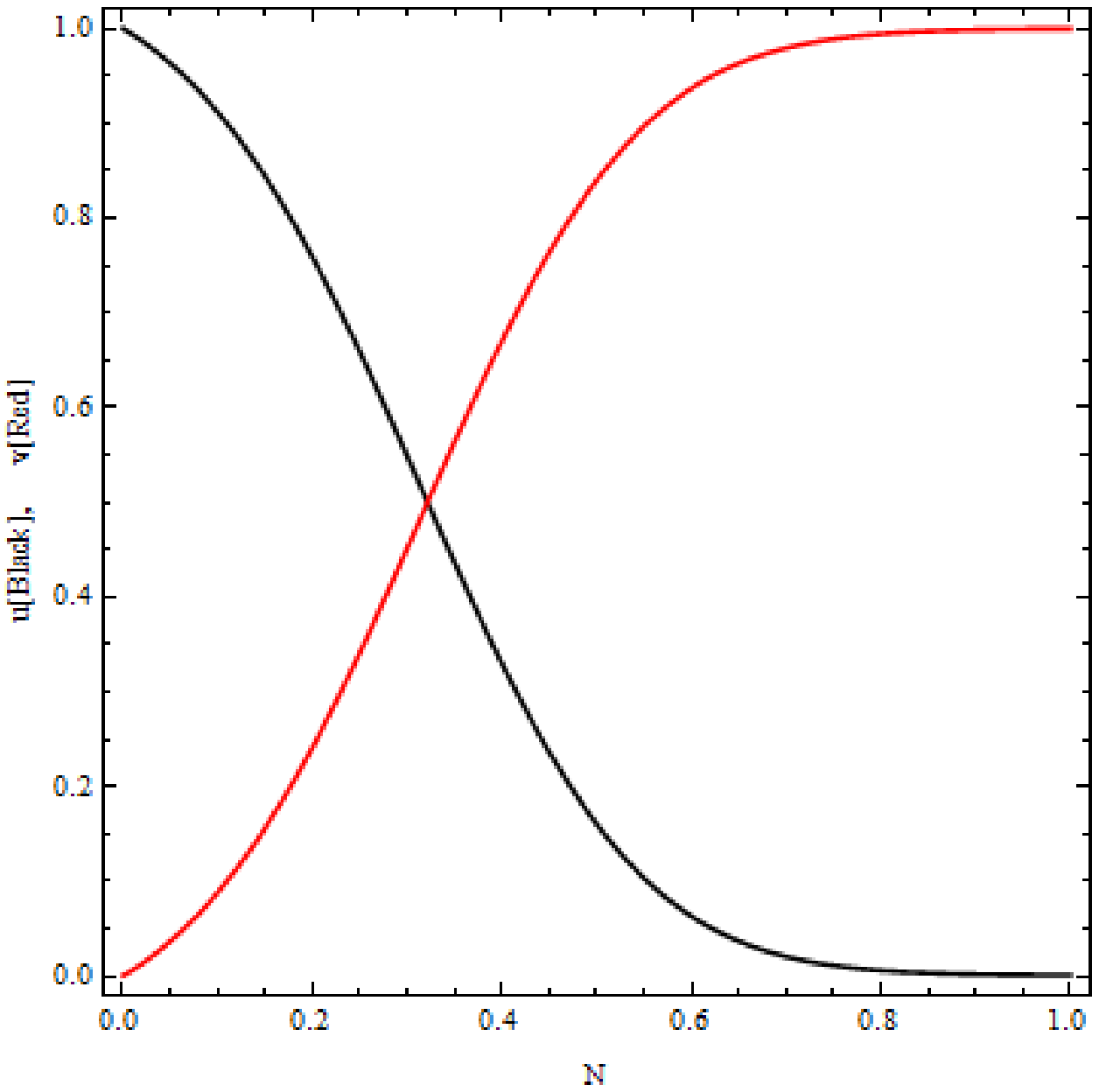} \label{e1}&
\includegraphics[height=7cm,width=7cm]{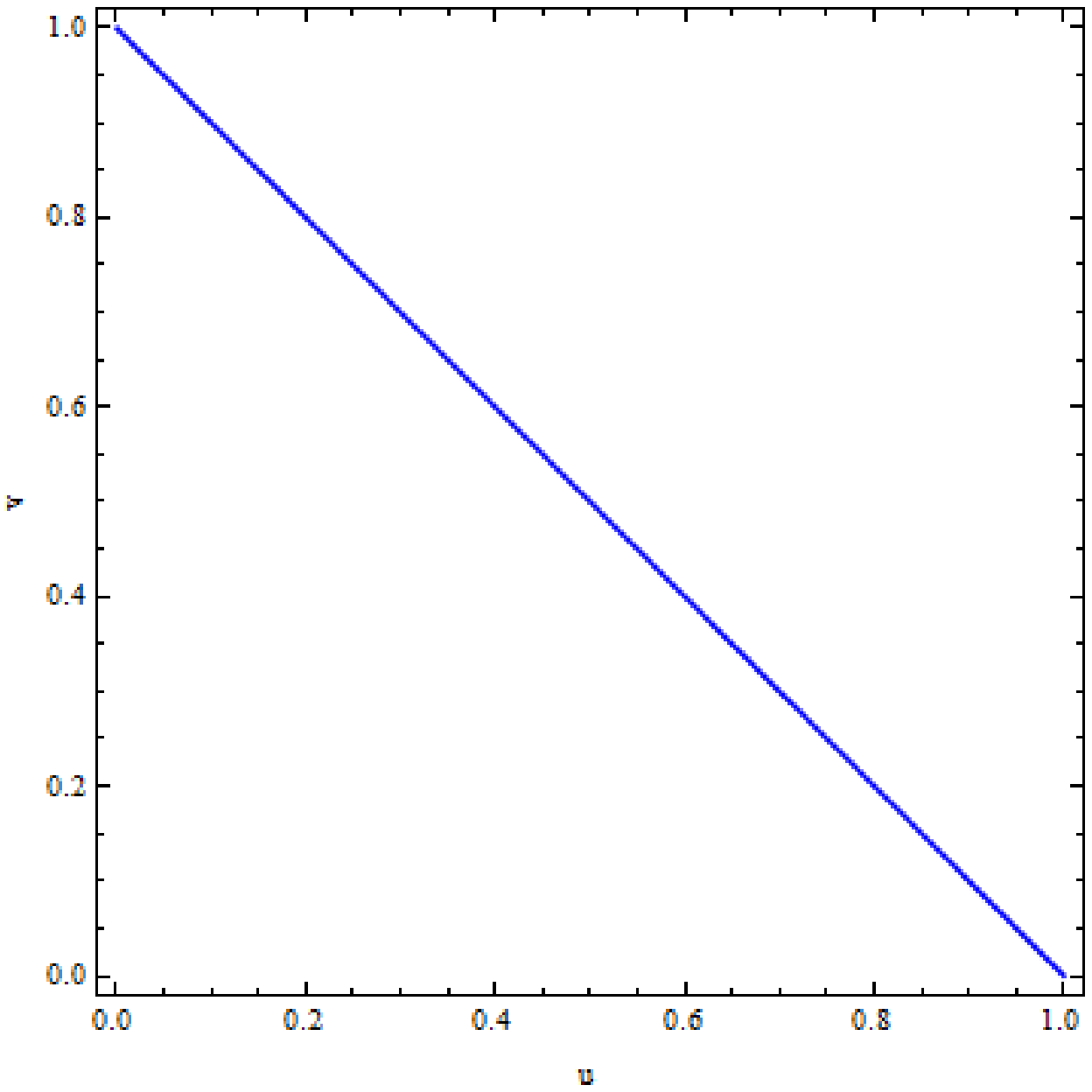} \label{e2}
\end{tabular}
\caption{ The graph at the left hand of this figure shows the evolution of the cosmological  parameters $u$ 
and $v$ versus the $e$-folding parameter $N$. The graph at the right hand shows the space-phase, $v$ versus $u$.
The graphs are plotted for the interaction term $q=0.2$, and the parameter of equation of state of ordinary content of
the universe $w=0$, and the one characterising the dark energy $\omega_{d}=-2.9$.}
\label{fig1}
\end{figure}

We consider the universe is filled by two interacting fluids, the dark energy and the ordinary matter whose energy density  are respectively $\rho_d$ and $\rho$. This means that the energy corresponding to each fluid is not conserved and the semi-continuity equations of continuity are written as 
\begin{eqnarray}
\dot{\rho}+3H(\rho+p)= -E_1, \label{1}\\
\dot{\rho_{d}}+3H(\rho_{d}+p_{d})= E_1, \label{2}
\end{eqnarray}
where $ E_1$  denotes the term of interaction between the two fluids. Let us define the following cosmological density parameters
\begin{eqnarray}
u= \frac{\rho}{3H^{2}},\quad v= \frac{\rho_d}{3H^{2}}.
\label{a6}
\end{eqnarray}
By using the  $e$-folding parameter $ N = \ln{a}$, $a$ being the scale factor, the equations of continuity  $(\ref{1})$ and  $(\ref{2})$ are presented as a system:
\begin{eqnarray}
\left\{\begin{array}{rl}
\frac{du}{dN}= 3u[(u-1)(1+w)+v(1+w_{d})]- \frac{E_1}{3H^3}, \label{3}\\
\frac{dv}{dN}= 3v[(v-1)(1+w_{d})+v(1+w)]+ \frac{E_1}{3H^3} \label{4}\,,
\end{array}
\right.
\end{eqnarray}
where we assumed  $ \kappa^{2} = 1$. The critical point are found by setting 
\begin{eqnarray} 
\frac{du}{dN}=0,\quad \frac{dv}{dN} = 0\,.
\end{eqnarray} 
Considering that the interaction term $ E_{1} = 3qH\rho$ with $q $ a constant, one gets 
\begin{eqnarray}
\left\{\begin{array}{rl}
 3u[(u-1)(1+w)+v(1+w_{d})]-3qu = 0 ,
 \label{a6}\\
 3v[(v-1)(1+w_{d})+u(1+w)]+ 3qu = 0 .
 \label{a7}
 \end{array}
\right.
\end{eqnarray}
After a resolution we find the following critical points 
\begin{eqnarray}
 A(  u_{c}= 0, v_{c} = 1),  \quad B( u_{c}= \frac{(q+w-w_{d})}{w- w_{d}}, v_{c}= \frac{-q}{w-w_{d}}).
\end{eqnarray}
The critical point $B$ seems reasonable and it is about it we will perform the study of stability. Then, we consider a perturbation in the vicinity of this point and write the variable $u$ and $v$ as
\begin{eqnarray}
u= u_{c}+\delta{u},\quad  v = v_{c}+\delta{v}.
\end{eqnarray}   
Therefore, the system becomes 
\begin{eqnarray}
\left\{\begin{array}{rl}
\frac{d\delta{u}}{dN}=3[(2u_{c}-1)(1+w)+(1+w_{d})v_{c}-q]\delta{u}+3u_{c}(1+w_{d})\delta{v},\label{5}\\
\frac{d\delta{v}}{dN}=3[v_{c}(1+w+q)\delta{u}+3[(2v_{c}-1)(1+w_{d})+(1+w)u_{c}]\delta{v},\label{6}
\end{array}
\right.
\end{eqnarray}
Regarding the critical point $B$ the eigenvalues are found as 
\begin{eqnarray}
\lambda_{1}= 3(q+w-w_{d}),\quad \lambda_{2}= 3(1+q+w).
\end{eqnarray} 
It is easy to see that for 
 $w_{d} < w+q $ and  $ -1 < w+q $, $\lambda_{1}$ and  $\lambda_{2}$ are positives. Hence, the point $B$ is unstable.  For  $\lambda_{1}\lambda_{2}$  $ < 0 $, $w_{d} < w+q < -1 $ or $ -1 < w+q < w_{d}$, $B$ is saddle point. For $\lambda_{1}< 0$, $\lambda_{2}< 0$, $q+w <  w_{d}$ and $ q+w < - 1$, $B$ is stable (an attractor).
We present the evolution of the variable $u$ and $v$ versus the $e-$folding parameter $N$ as well as the space phase for some 
suitable values of the input parameters consistent with the observational data, and presented at Fig. \ref{fig1}. We see from this figure that the interaction case, as the energy density of the ordinary matter decreases and becomes very small, the one of the dark energy goes increases and goes toward $1$ as the time evolves. This is quite reasonable because from observational data, the current universe is well dominated by the dark energy. On the other hand, regarding the graph of $v$ versus $u$, it is clear that the critical point about which the stability is studies is a saddle point.

\section{Stability of $f(R,T)$ model}\label{section4}

This section is devoted to the study of the stability of the model $ f(R, T)= R+f(T) $  using the power law and de Sitter solution.\par 
We will be interested to the perturbation of both the geometrical and matter parts of the generalized equations of motion. To do so, we focus our attention to the Hubble parameter for what concerns the geometry and the energy density of the ordinary content concerning the matter of the background, and perform the perturbation about them as \cite{antoniode,diegoalvaro}
\begin{eqnarray} 
 H(t) = H_{b}(t)(1+\delta(t)),\quad   \rho(t)= \rho_{b}(t) (1+\delta_{m}(t))\label{7} ,
\end{eqnarray}  
where $H_b(t)$ and $\rho_b(t)$ denote the Hubble parameter and the energy density of the ordinary matter of the background respectively. Taking into account the interaction term, the continuity equation of the ordinary matter is cast into the form 
\begin{eqnarray}    
\dot{\rho_{b}}(t)+3H_{b}(t)\rho_{b}(t)(1+w+q)= 0\,, \label{8}
\end{eqnarray}        
which is solved giving 
\begin{eqnarray}
\rho_{b}(t)=\rho_{0}\ e^{-3(1+w+q)\int {H_{b}(t)}dt},\label{9}
\end{eqnarray}
where $\rho_{0}$ is an integration constant. In order to study the linear perturbation about $ H(t)$ and  $ \rho(t)$, we develop  $f(T)$ in a series of $T_{b} = \rho_{b}(1-3w)$ as:
\begin{eqnarray}
f(T) = f^{b}+f_{T}^{b}(T-T_b) + O^{2}\label{10}, 
\end{eqnarray}
The function $ f(T) $  and its derivative are evaluated at $T=T_{h}$. Regarding the Einstein-Hilbert term, the novelty here is the effect coming from $f(T)$. By substituting (\ref{7}) and (\ref{10}) into the first generalized equation of Friedmann,
\begin{eqnarray}
3H^{2} = \rho+ (\rho+p)f_{T} + \frac{1}{2}f(T)\,,
\end{eqnarray}
one gets after simplification 
\begin{eqnarray}
6H_{b}^{2}(t)\delta(t)= \big[\rho_{b}+\rho_{b}f_{T}^{b}(\frac{3-w}{2})+\rho_{b}^{2}(1-2w-3w^{2})f_{TT}^{b}\big]\delta_{m}(t).
\label{11}
\end{eqnarray}
Considering that the ordinary matter is essentially the dust, we obtain the simple expression 
\begin{eqnarray}
6H_{b}^{2}(t)\delta(t)= \big[\rho_{b}+3\rho_{b}f_{T}^{b}+2\rho_{b}^{2}f_{TT}^{b}\big]\delta_{m}(t)\label{12}.
\end{eqnarray}
Regarding the matter perturbation function one gets the following differential equation 
\begin{eqnarray}
\dot{\delta_{m}}(t)+3(1+w+q)H_{b}(t)\delta(t) = 0. \label{13}
\end{eqnarray}
Eliminating $\delta(t)$ between (\ref{11}) and (\ref{13}), we obtain the differential equation 
\begin{eqnarray}
2H_{b}\dot{\delta}_{m}(t)+(1+w+q)
\big[\rho_{b}+\rho_{b}f_{T}^{b}\left(\frac{3-w}{2}\right)+\rho_{b}^{2}(1-2w-3w^{2})f_{TT}^{b}\big]\delta_{m}(t)=0,\label{14}
\end{eqnarray} 
whose general solution reads 
\begin{eqnarray}
\delta_{m}(t)= C_{0}\exp\left\{-\left(\frac{1+w+q}{2}\right)\int  \frac{\rho_{b}}{H_{b}}\left[1+f_{T}^{b}\left(\frac{3-w}{2}\right)+\rho_{b}(1-2w-3w^{2})f_{TT}^{b}\right]dt \right\} ,
\label{15} 
\end{eqnarray}
where $C_{0}$ is an integration constant. From Eq.~(\ref{13}) one extract 
\begin{eqnarray}
\delta(t) = \frac{C_{0} C_{T}}{6H_{b}}\exp\left\{ -\left(\frac{1+w+q}{2}\right)\int C_{T}dt \right\}  ,
\label{16}
\end{eqnarray}
with 
\begin{eqnarray}
C_{T}=\frac{\rho_{b}}{H_{b}}\left[1+f_{T}^{b}\left(\frac{3-w}{2}\right)+\rho_{b}\left(1-2w-3w^{2}\right)f_{TT}^{b}\right].
\end{eqnarray}

\subsection{Stability of de Sitter solutions}

In this case, the Hubble parameter is written as  
\begin{eqnarray}
H_{b}(t) = H_{0} \rightarrow a(t) = a_{0}e^{H_{0}t}.
\end{eqnarray}
The expression (\ref{9}) becomes,
\begin{eqnarray}
\rho_{b}(t)=\rho_{0} e^{-3(1+w+q)H_{0}t}.
\label{17}
\end{eqnarray}
Making use of the relation $ d{\rho_{b}}=-3(1+w+q)H_{0}\rho_{b}dt$ and through an elementary transformation one gets 
\begin{eqnarray}
\int C_{T}dt &=&-\frac{1}{3H_{0}(1+w+q)}\int\frac{1}{\rho_{b}}C_{T}d{\rho_{b}} \nonumber\\
&=&-\frac{1}{3H_{0}^{2}(1+q+w)}\Bigg\{ \rho_{b}
+\alpha \frac{(3-w)}{2}    \rho_{b}^{\beta}(1-3w)^{\beta-1} + \alpha(1-2w-3w^{2}) (\beta-1)(1-3w)^{\beta-2}\rho_{b}^{\beta} \Bigg\}.
\end{eqnarray}
By replacing this expression in  (\ref{15}), one  obtains 
\begin{eqnarray}
\delta_{m}(t)= C_{0}\exp \left\{ \frac{1}{6H_{0}^{2}}\left[\rho_{b}+\alpha \frac{(3-w)}{2}    \rho_{b}^{\beta}(1-3w)^{\beta-1} + \alpha(1-2w-3w^{2}) (\beta-1)(1-3w)^{\beta-2}\rho_{b}^{\beta}\right]\right\}.
\label{18}
\end{eqnarray}
Therefore the perturbation function about the geometry can be obtained, given by 
\begin{eqnarray}
\delta(t) = \frac{C_{0} C_{T}}{6H_{0}}\exp \left\{ \frac{1}{6H_{0}^{2}}\left[\rho_{b}+\alpha \frac{(3-w)}{2}    \rho_{b}^{\beta}(1-3w)^{\beta-1} + \alpha(1-2w-3w^{2}) (\beta-1)(1-3w)^{\beta-2}\rho_{b}^{\beta}\right]\right\},
\label{19}
\end{eqnarray}
with
\begin{eqnarray}
 C_{T}= \frac{1}{H_{0}}\left\{\rho_{b}+\alpha\beta \frac{(3-w)}{2}    \rho_{b}^{\beta}(1-3w)^{\beta-1} + \alpha\beta  (\beta-1)(1-2w-3w^{2})(1-3w)^{\beta-2}\rho_{b}^{\beta}\right\},
\end{eqnarray}
and 
\begin{eqnarray}
 f(T) = \alpha T^{\beta}, 
\end{eqnarray}
 with  $\alpha$ the one defined in (\ref{eti15}) and $\beta=(1+3\omega)/(2(1+\omega))$.
For some suitable values of the input parameters consistent with the cosmological observational data, we plot the curve characterizing the behaviour of the perturbation function at the left side in Fig. \ref{fig2}. We see that as the universe expands, i.e., increasing $N$, the matter and geometric perturbations functions, $\delta_m$ and $\delta$ respectively, goes towards positive values more less than $0.1$ as the time evolves.

\subsection{Stability of Power-Law solutions}

\begin{figure}[h]
\centering
\begin{tabular}{rl}
\includegraphics[height=7cm,width=7cm]{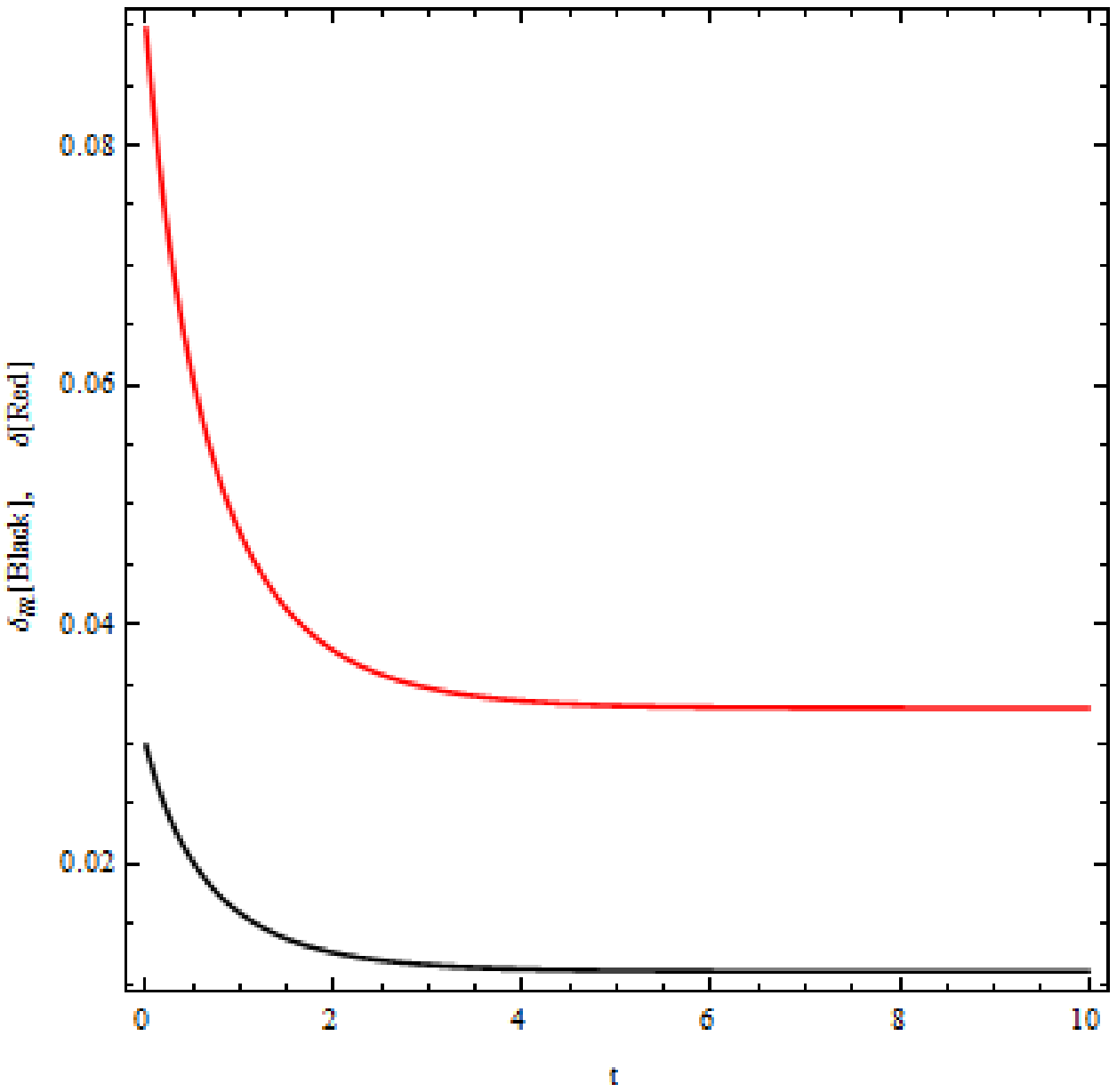} \label{e3}&
\includegraphics[height=7cm,width=7cm]{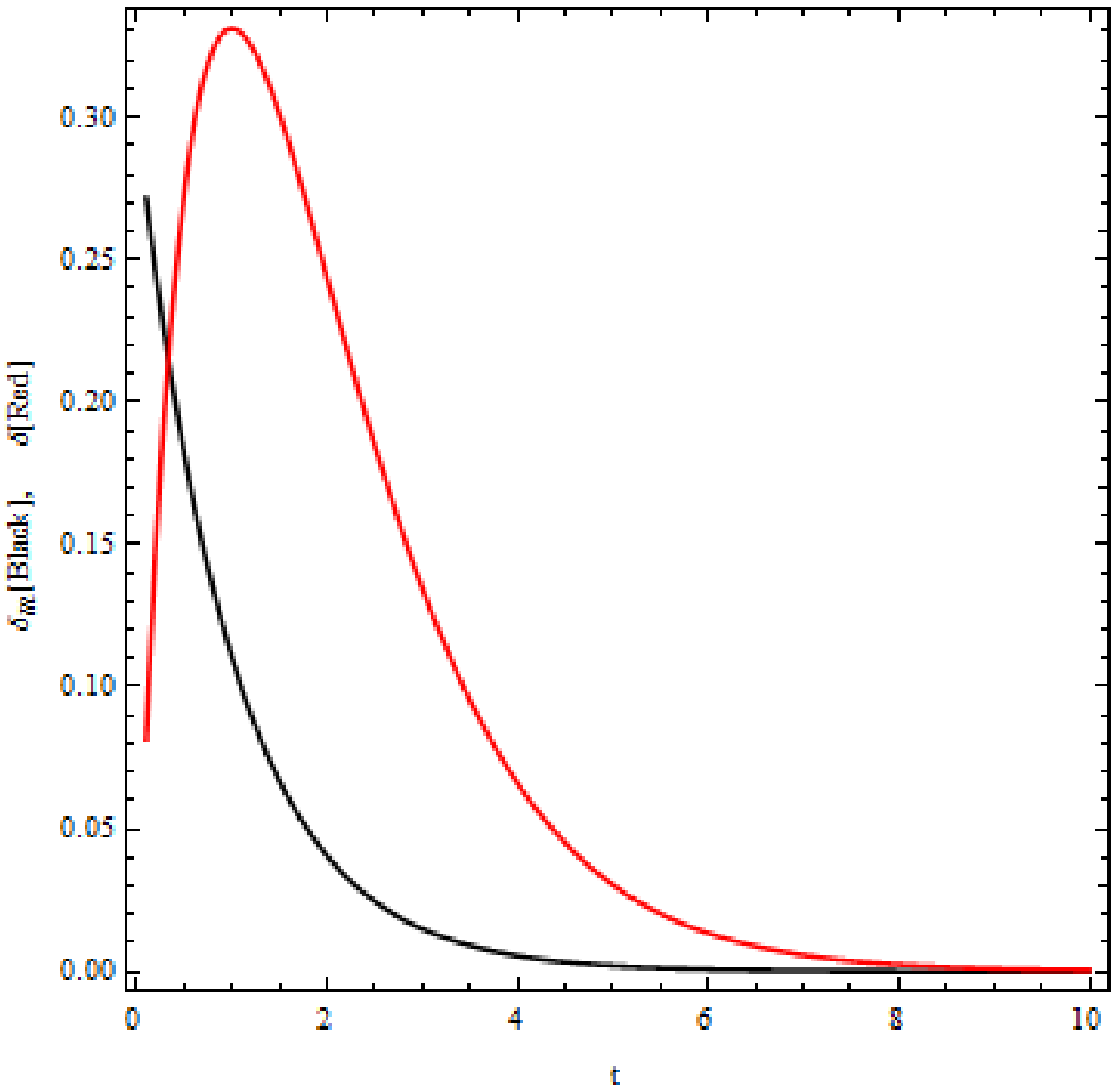} \label{e4}
\end{tabular}
\caption{ The graph at the left side of the figure presents the evolution of the perturbation functions $\delta_m$ (black) 
and $\delta$ (red) within the de Sitter solutions, while the one at the right side shows the evolution of the perturbation
functions within the power-law solutions. The graph are plotted for $n=2/3$, $\Lambda=1.7\times 10^{-121}$, $\rho_0=0.1\times 10^{-121}$, 
$\omega=0$ and $C_1=1$ .}
\label{fig2}
\end{figure}

Here, the scale factor is written as  
\begin{eqnarray}
 a(t)\propto t^{n} \quad \rightarrow   H_{b}(t) = \frac{n}{t}\,,
\end{eqnarray} 
 and the ordinary energy density (\ref{9}) becomes
\begin{eqnarray}
\rho_{b}=\rho_{0}t^{-3n(1+w+q)}\label{20}.
\end{eqnarray}
By making the substitution of  $\rho_{b}$ in  (\ref{14}), one gets after resolution, the following expression
\begin{eqnarray}
\delta_{m}(t)= C_{1}\exp\left\{ -A\left(\frac{A_{1}}{2+B}t^{2+B} +\frac{A_{2}}{2+\beta B}t^{2+\beta B}\right)\right\},
\end{eqnarray}
 where $C_{1}$ is an integration constant, and 
\begin{eqnarray} 
A = \frac{(1+w+q)}{2n}, \quad A_{1}= \rho_{0},\quad B = -3n(1+w+q),\nonumber\\ A_{2}=\alpha\beta \rho_{0}^{\beta}\left\{ \frac{(18w^{3}+9w^{2}-14w+3)}
{2(1-3w)^{2-\beta}}+\frac{\beta}{{(1-3w)}^{(1-\beta)}}\right\}.
\end{eqnarray} 
 From the expression (\ref{13}), one obtains 
\begin{eqnarray} 
\delta(t)=\frac{C_{1}}{6n^{2}}\left( A_{1}t^{2+\beta}+A_{2} t^{2+B \beta}\right) \exp\left\{ -A\left(\frac{A_{1}}{2+B}t^{2+B} +\frac{A_{2}}{2+\beta B}t^{2+\beta B}\right)\right\}.
 \end{eqnarray}
 As performed in the previous subsection, we present the evolution of the perturbation functions for suitable cosmological values 
 of the input parameters in Fig \ref{fig2}.

\section{Cosmological dynamics in $R+f(T)$ gravity}\label{section5}

This rubric is devoted to the study of the model of type  $ f(R,T) = R +f(T)$ using the cosmological solutions of  
Low-redshift and  high-redshift. Here, we decouple the matter in its relativist part (the radiation) and  its non-relativist 
(assumed as the dust), then, assuming that the interaction occurs between the dust and the dark energy.\par
The equations of continuity of the considered fluids are written as  
\begin{eqnarray}
\left\{\begin{array}{rl}
\dot{\rho}_{m}+3H\rho_{m}\left(1+q\right) = 0\,,\\
\dot{\rho}_{rad}+4H\rho_{rad}= 0\,,
\\
\dot{\rho}_{d}+3H\rho_{d}
\left(1+\omega_{d}\right)=3Hq\rho_m,
\end{array}
\right.
\end{eqnarray}  
where we have assumed that the interaction term between the dark energy and the ordinary matter is $3Hq\rho_m$.

\subsection{Low-redshift solutions} 

In order to develop the study it is suitable to  introduce the quantities 
\begin{eqnarray}
 x\equiv \frac{\dot{H}}{H^{2}},\quad  y \equiv H,\quad \Omega_{m}
 \equiv \frac{\rho_{m}}{3H^{2}},\quad\Omega_{rad}
 \equiv \frac{\rho_{rad}}{3H^{2}}.
\end{eqnarray} 
Hence, using the first generalized equation of Friedmann, one gets the following system
\begin{eqnarray}  
\left\{\begin{array}{rl}
x'  = \frac{9}{2}(1+q)(1+f_{T}+f_{TT})y \Omega_{m}+8(1+f_{T}+\frac{3}{4} f_{TT})y\Omega_{rad}-2x^{2},\\
y'= xy,\\
\Omega'_{m} = -(3+2x)\Omega_{m},\\
\Omega'_{rad}= -(4+2x)\Omega_{rad},
\end{array}
\right.
\end{eqnarray}
where the prime denotes the derivative with respect to the parameter  $N=\ln{a}$. Once our model is specified, we can integrate 
the background equations through the above dynamic system directly in the low-redshift regime, within some suitable cosmological values of
the input parameters. Through the well known relations $\Omega_{DE}=1-\Omega_m-\Omega_{rad}$ and $\omega_{eff}=-1-2\dot{H}/(3H^2)=-1-2x/3$,
we present the evolution of $\omega_{eff}$, $\Omega_{DE}$, $\Omega_m$ and $\Omega_{rad}$ versus the $e-$folding parameter. The initial conditions
are assumed to be $x=-1.502$, $y=20.0$, $\Omega_m=0.9959$ and $\Omega_{rad}=0.004$. This numerical results show clearly the cosmological
evolution of the universe, i.e., the matter era is followed by the accelerated one with final stage producing exactly the $\Lambda CDM$ 
feature where $\omega_{eff}$ goes toward $-1$ for large scalar factor (low-redshift). \par Moreover, we try to check the influence of the 
interaction term by first vanishing it ($q=0$). In such a 
situation, comparing with the case where the interaction term is considered, we see that the effective parameter of equation of
state $\omega_{eff}$ is more negative. This is quite reasonable and explain why is it important to consider a running cosmological 
constant to realize the interaction between the matter and dark energy. Observe that when $q=0$, $\omega_{eff}$ is more negative and the 
universe will be more phantom than in the case where interaction term is considered. Hence, it appears clearly that in order to have a
$\omega_{eff}$ consistent with the observation data (say,$ WMAP 9$ data \cite{bennett}), interaction terms are indispensable. The model
under consideration in this paper offers this possibility of more reaching the physical interval of $\omega_{eff}$ according to 
observational data. Therefore, the model is acceptable, at least in the view of low-redshift regime, for being a competitive candidate
for the dark energy.
 \begin{figure}[h]
\centering
\begin{tabular}{rl}
\includegraphics[height=6cm,width=6cm]{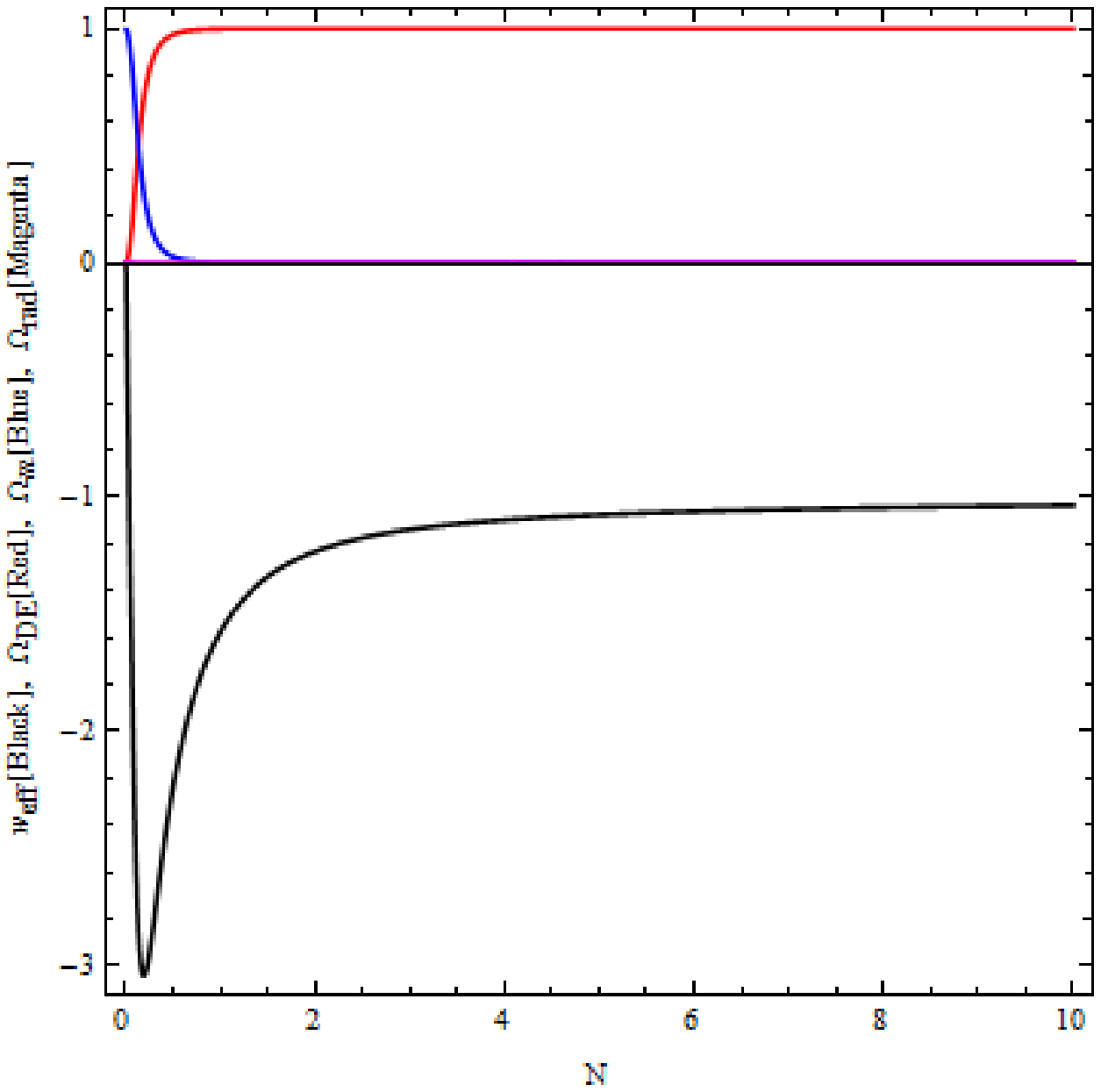}\label{e5} &
\includegraphics[height=6cm,width=6cm]{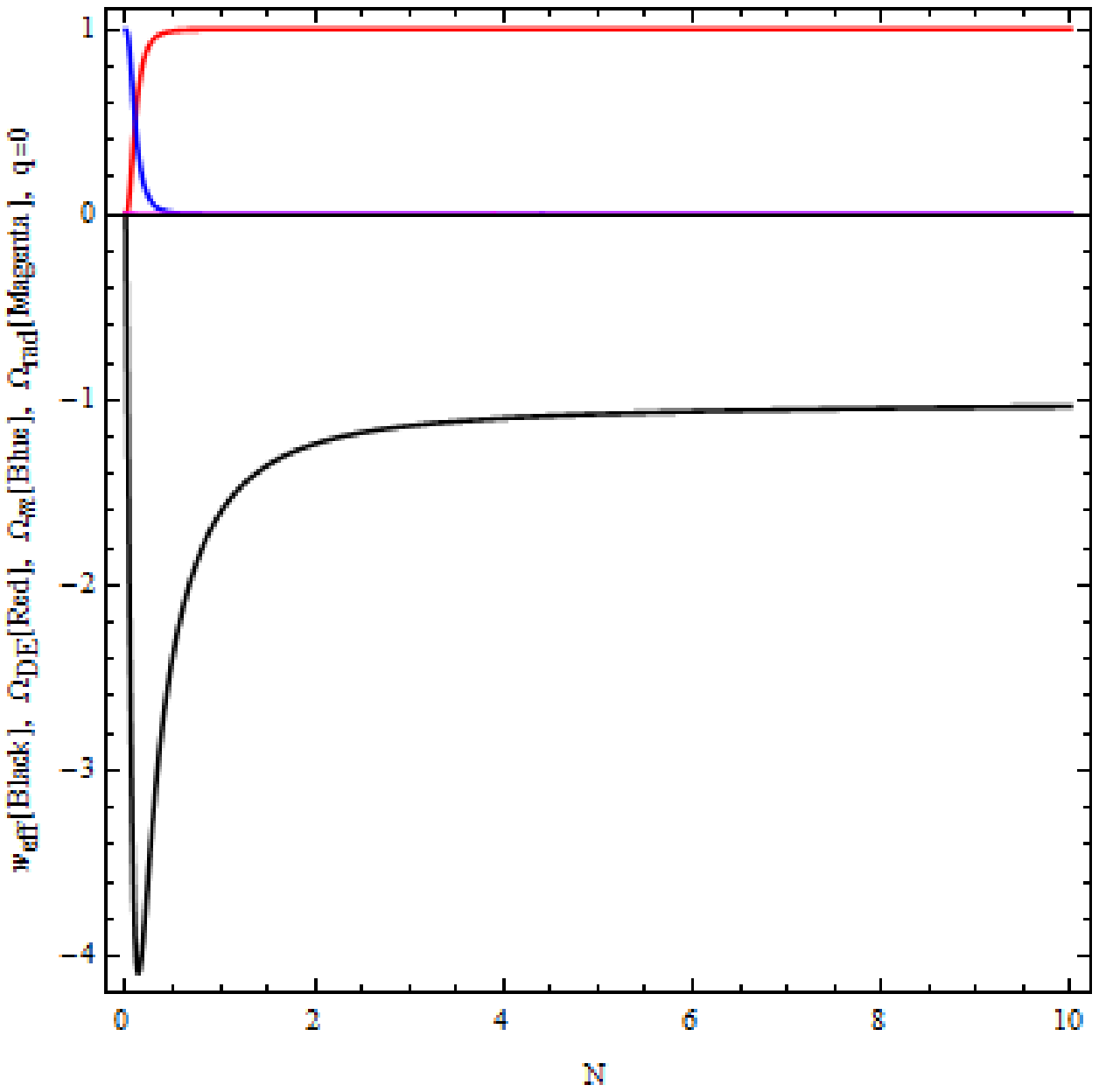}\label{e6}\\
\includegraphics[height=6cm,width=6cm]{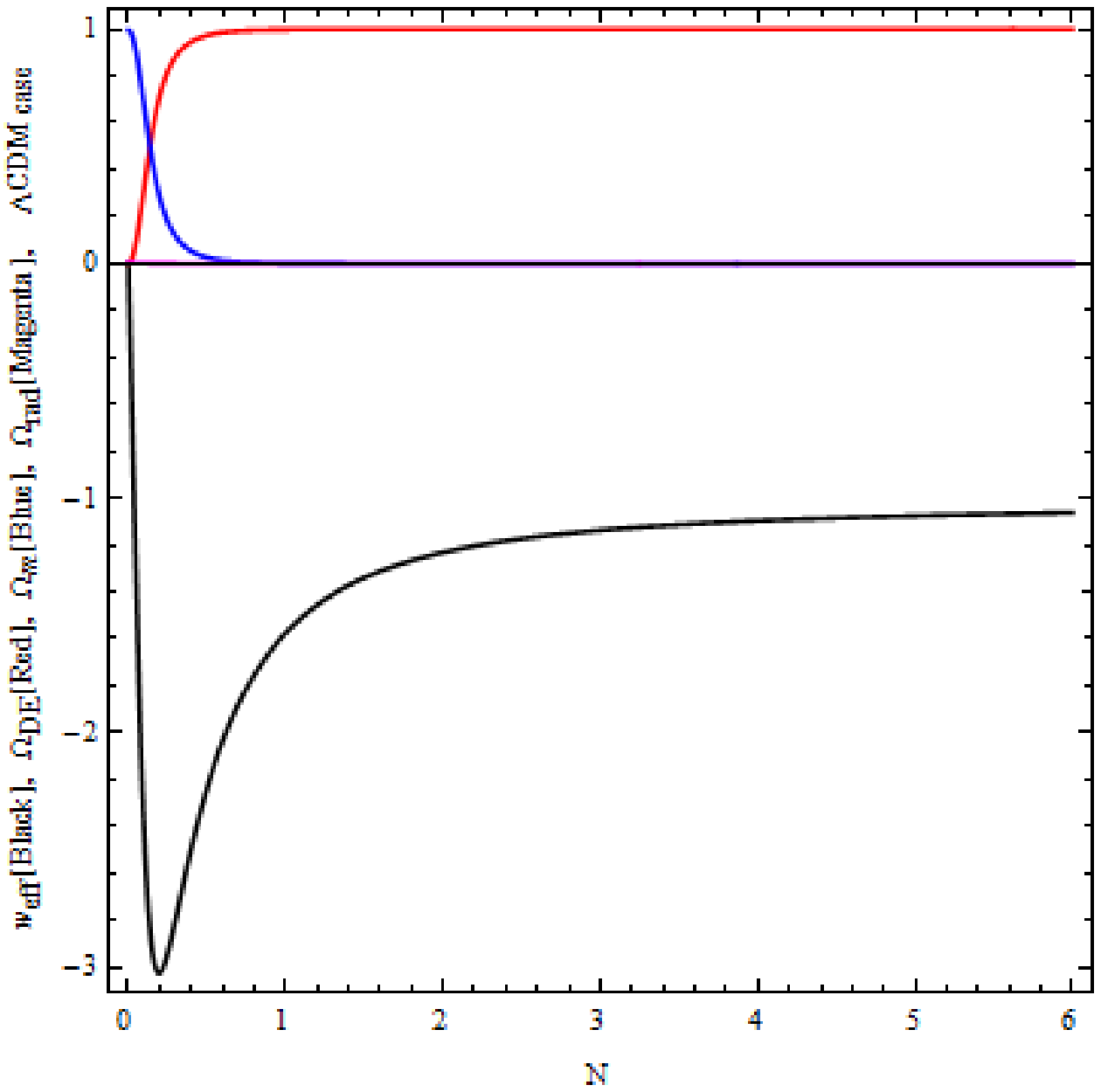}\label{e7}
\end{tabular}
\caption{ The graphs present the evolution of the cosmological parameters $\Omega_m$ (blue), $\Omega_{rad}$ (magenta),
$\Omega_{DE}$ (red), and $\omega_{eff}$ (black). The left one describes the evolution of the above parameters
versus $N$ within the model under consideration with interaction term   ($q=-0.2$), while the middle one corresponds to the 
case where the interaction term vanishes ($q=0$). The right one describes the behaviour of the parameters in $\Lambda CDM$ model with ($q=-0.2$). 
The graphs are plotted for $\omega=0$, $\rho_0=0.1\times 10^{-121}$ and $\Lambda CDM = 1.7\times 10^{-121}$.}
\label{fig3}
\end{figure} 

\subsection{High-redshift solutions} 

In this subsection  we search for the viability of the model for very small values of the scale corresponding to high-redshift, $a\lesssim 10^{-2}$, (the era of nucleosynthesis). In such a situation, one can minutely fix the initial value of $N$ at that for which $a=10^{-2}$, that is $N\simeq -4.6$. Let us illustrate this by first writing the first generalized Friedman equation as
\begin{eqnarray}
H^2-H_{\Lambda}^2=\frac{1}{3}\left[\frac{1+\omega}{1-3\omega}Tf_T+\frac{1}{2}f\right],\label{eti52}
\end{eqnarray}
where
\begin{eqnarray}
\frac{H^2_{\Lambda}}{H_0^2}=\frac{\Omega_m^{(0)}}{a^3}+\frac{\Omega_{rad}^{(0)}}{a^4}\,\,.
\end{eqnarray}
It is easy to see that for times close to the current one, that is, $T \rightarrow T_0$, the algebraic function 
$f(T)$ goes toward $-2\Lambda$ such that $f_T\rightarrow 0$   and the right hand side of (\ref{eti52}) gives $-\Lambda/3$. Now, at 
high-redshift, i.e., for small scale factor, so, negative $N$, one can explore the evolutions of the same parameters as performed in 
the above case (the low-redshift). At the same time, we need to be more realist taking into account the physical aspect of the universe.
It is important to note that for high-redshift, the universe lives in Plank era when the universe is practically filled by radiation.
Considering that all this happens in the first Planck time, where the nucleo-synthesis is starting, we can set the initial conditions 
to $0.004$ for the matter and $0.9959$. As it is well known, as the universe expands, it evolves becoming cool and the $quark$ known as
the elementary particle gives rive to all matter we see today through the recombination process. If $\
\Lambda CDM$ is widely accepted to describe to universe, mainly at it present stage, we try to go toward early times and analyse
the deviation of the model under consideration in this paper, with the $\Lambda CDM$ one. To do so, we plot the right hand side 
of (\ref{eti52}) and compare it numerically with the $\Lambda CDM$ for high-redshift. The result is presented at the Fig. $4$. This 
result shows that the model under consideration is this paper reproduces the $\Lambda$ behaviour for small scale factor. Indeed, this 
result was 
expected because for high-redshift, the radiation is predominant and the effective parameter of equation of state $\omega_{eff}$ is 
about $1/3$. In such a situation, the trace $T$ of the energy momentum tensor vanishes and the $\Lambda CDM$ model should be recovered.
Therefore, the model also works very for the early moment of the universe. However, as the universe evolves, the $f(R,T)$ model starts 
differing slightly from the $\Lambda CDM$ one. 

\begin{figure}[htbp]
\begin{center}
\includegraphics[width=9cm, height=9cm]{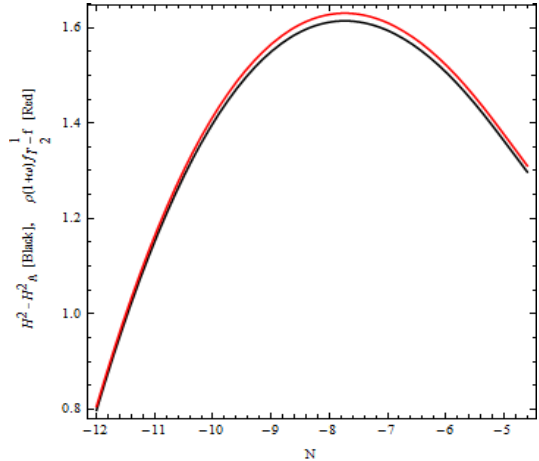}
\end{center}
\caption{The graph shows the deviation of the $f(T)$ from 
$\Lambda CDM$ model. The graph are plotted for $\omega=1/3$ and $\Lambda CDM = 1.7\times 10^{-121}$ }
\label{fig4}
\end{figure}

\section{Conclusion}\label{section6}  

We undertook in this work cosmological analysis about a model in the framework of the so-called $f(R,T)$ theory. In order to obtain a viable $f(R,T)$ model, we first impose the covariant conservation of the energy momentum, from which, we get a model of the king $R+f(T)$, being a sort of trace depending function correction to the general relativity. The obtained model includes parameters depending on the cosmological constant $\Lambda$ and the parameter $\omega$ of the ordinary equation of state. These parameters play a main role in the whole study developed in this manuscript. By the way, we study the dynamics of the cosmological system, analysing the stability about the critical points. Our result shows
that for both de Sitter and power-law solutions, the perturbations functions converge traducing the stability of the model. \par
Moreover, the stability of the model is checked within the de Sitter and power-law solutions by performing linear perturbation about the physical critical point. We see that for the both considered  solutions, the model presents stability through the convergence of the geometric and matter perturbation functions $\delta$ and $\delta_m$.\par
Regarding the cosmological dynamics, we search how much the model may describe the early and present stages of the universe by studying how much it is consistent in both the low-redshift and high-redshift regimes. The results present consistency with the cosmological observational data.\par Therefore, we conclude that, regarding the stability and dynamics, the model under study here is competitive candidate for dark energy.

\vspace{1.5cm}

{\bf Acknowledgement}: The authors thank Prof. S. D. Odintsov for useful comments and suggestions. E. H. Baffou  thanks IMSP for every kind of support during the realization of this work.


\begin{thebibliography}{17}

\addcontentsline{toc}{chapter}{Bibliographie}

\bibitem{1dediego}
 S. Nojiri and S. D. Odintsov, eConf \textbf{C0602061}, 06 (2006) [Int. J. Geom. Meth. Mod. Phys. \textbf{4}, 115 (2007)]; hep-th/0601213; arXiv: 0807.0685; Phys. Rept {\bf 505}, 59-144 (2011). 
  K.~Bamba, S.~Capozziello, S.~Nojiri and S.~D.~Odintsov,
  arXiv:1205.3421 [gr-qc];
   A. De Felice and S. Tsujikawa, Living Rev.\ Rel.\  {\bf 13}, 3 (2010) [arXiv:1002.4928 [gr-qc]];
  V. Faraoni 
 arXiv:0810.2602v1 [gr-qc]; F. S. N. Lobo, arXiv: 0807.1640 [gr-qc];   S. Capozziello  and  V. Faraoni, {\it Beyond Einstein Gravity}, Fundamental Theories of Physics Vol. 170, Springer Ed., Dordrecht  (2011);
   S.~Capozziello and M.~Francaviglia,
  Gen.\ Rel.\ Grav.\  {\bf 40}, 357 (2008)
  arXiv:0706.1146 [astro-ph];
  S.~Capozziello and M.~De Laurentis,
  Phys.\ Rept.\  {\bf 509}, 167 (2011)
  [arXiv:1108.6266 [gr-qc]];
  A.~de la Cruz-Dombriz and D.~S\'aez-G\'omez,
  Entropy {\bf 14}, 1717 (2012)
  [arXiv:1207.2663 [gr-qc]].

\bibitem{2dediego} B. Boisseau, G. Esposito-Farese, D. Polarski and A. A. Starobinsky, Phys. Rev. Lett. {\bf 85}, 2236 (2000) [arXiv:gr-qc/0001066]; 
S. M. Carroll, I. Sawicki, A. Silvestri and M. Trodden, New J. Phys. {\bf 8}, 323 (2006)
[arXiv:astro-ph/0607458]; G. Esposito-Farese and D. Polarski, Phys. Rev. D {\bf 63}, 063504 (2001) [arXiv:gr-qc/0009034]; 
P. Zhang, Phys. Rev. D {\bf 73}, 123504 (2006) [arXiv:astro-ph/0511218].

\bibitem{lambdadeT} Nikodem J. Poplawski,  arXiv:gr-qc/0608031v2.

\bibitem{frtoriginal}
T. Harko, F. S. N. Lobo, S. Nojiri and S. D. Odintsov, Phys. Rev. {\bf D84} (2011) 024020. [arXiv:1104.2669 [gr-qc]]. 

\bibitem{lesfrt} M. J. S. Houndjo, Int. J. Mod. Phys. D. {\bf 21}, 1250003 (2012). arXiv: 1107.3887 [astro-ph.CO];
M. J. S. Houndjo and O. F. Piattella,  Int. J. Mod. Phys. D. {\bf 21},   1250024  (2012). arXiv: 1111.4275 [gr.qc];
D. Momeni, M. Jamil and R. Myrzakulov, Euro. Phys. J. C {\bf 72}, arXiv: 1107.5807[physics.gen-ph].

\bibitem{flavio} F. G. Alvarenga, M. J. S. Houndjo, A. V. Monwanou and Jean. B. Chabi-Orou, J. Mod. Phys., {\bf 4}, 130-139 (2013)
arXiv: 1205.4678 [gr-qc].

\bibitem{thermo1} M. Sharif and M. Zubair, JCAP {\bf 03}, 028 (2012); arXiv:1204.0848v2 [gr-qc].
  M.~Jamil, D.~Momeni and R.~Myrzakulov,
  Chin.\ Phys.\ Lett.\  {\bf 29}, 109801 (2012)
  [arXiv:1209.2916 [physics.gen-ph]].

\bibitem{juliano} M. J. S. Houndjo, C. E. M. Batista, J. P. Campos and O. F. Piattella, 
 Can. J. Phys. {\bf 91}, 548-553 (2013). arXiv:1203.6084 [gr-qc].



\bibitem{tahereh} Tahereh Azizi,  Int. J. Theor. Phys. {\bf 52}, 3486-3493 (2013). arXiv:1205.6957 [gr-qc].

\bibitem{farasat} 	M. Farasat Shamir, Adil Jhangeer, Akhlaq Ahmad Bhatti, arXiv:1207.0708 [gr-qc].

\bibitem{sharifzubair} 	Muhammad Sharif, Muhammad Zubair, J. Phys. Soc. Jap. {\bf 82}, 014002 (2013).  arXiv:1210.3878 [gr-qc]. 

\bibitem{chakra} Subenoy Chakraborty,  Gen. Rel. Grav. (2013), DOI: 10.1007/s10714-013-1577-y.  arXiv:1212.3050 [physics.gen-ph].

\bibitem{hamid} Hamid Shabani, Mehrdad Farhoudi,  Phys. Rev. D {\bf 88} 044048 (2013). arXiv:1306.3164 [gr-qc].

\bibitem{samanta} 	G. C. Samanta. Int. J. Theor. Phys. {\bf 52}, 2303-2315 (2013).  

\bibitem{santos} A.F. Santos, Mod. Phys. Lett. A {\bf 28}, 1350141 (2013). arXiv:1308.3503 [gr-qc].

\bibitem{2foismuhammad} Muhammad Sharif and Muhammad Zubair,  J. Phys. Soc. Jap. {\bf 82}, 064001 (2013).  arXiv:1310.1067 [gr-qc].

\bibitem{anil} Anil Kumar Yadav, arXiv:1311.5885 [physics.gen-ph].
  
\bibitem{papierdiego} F. G. Alvarenga, A. de la Cruz-Dombriz, M. J. S. Houndjo, M. E. Rodrigues, D. S\'aez-G\'omez,  Phys. Rev. D {\bf 87}, 103526 (2013).  arXiv:1302.1866 [gr-qc].

\bibitem{antoniode} A. De Felice and S. Tsujikawa, Phys. Lett. B {\bf 675}, 1-8 (2009); arXiv: 0810.5712 [hep-th].


\bibitem{diegoalvaro} A. de la Cruz-Dombriz and D. S\'aez-G\'omez, Class. Quantum Grav {\bf 29}, 245014 (2012), arXiv: 1112.4481 [gr-qc].

 \bibitem{bennett} C. L. Bennett {\it et al},  Accepted to Astrophysical Journal Supplement Series, arXiv:  arXiv:1212.5225v3 [astro-ph.CO]. 
  
\end{thebibliography}
\end{document}